\documentclass[fleqn,twoside,twocolumn,nofootinbib,showkeys]{revtex4} % Specifies the document class %,unsortedaddress
\usepackage[sec,nocpr]{ujp} % \usepackage[]{ujp} for cyrillic, \usepackage[web]{ujp} for web

\begin{document}
\title[FIRST-ORDER DIFFERENTIAL EQUATIONS]%колонтитул
{FIRST-ORDER DIFFERENTIAL EQUATIONS  \\ FOR A PARTICLE WITH SPIN \boldmath$S = 1$\,\,\,$^{1\,)}$}%

\author{B.E.~Grinyuk}%1 автор
\affiliation{\bitp}%институт
\address{\bitpaddr}%адрес
\email{bgrinyuk@bitp.kiev.ua
}%e-mail

\udk{530.145 + 539.12.01} \pacs{\\03.65.Pm, 11.10.-z,
\\[-3pt]11.10.Ef
}
\razd{\seci}

\autorcol{B.E.~Grinyuk}

\setcounter{page}{1447}%

\begin{abstract}
A system of first-order differential equations for a particle with
nonzero mass and spin $S = 1$ is constructed. As distinct from the
Proca--Duffin--Kemmer (PDK) equations, the system has the form of
the dynamical equation\,
$i\hbar\partial_{t}\hat{\Psi}=\hat{\mathrm{H}}\hat{\Psi}$ (with
constraints) with a Hamiltonian linear in momentum. The
six-component wave function yields the positive-definite
probability density $\hat{\Psi}^{\dag}\hat{\Psi}\geq 0$. The
system of equations has much in common with the Dirac and Maxwell
equations.
\end{abstract}


\maketitle

\section{Introduction}
The explicitly covariant PDK equations for a particle with spin $S
= 1$ cannot be presented in the form
$i\hbar\partial_{t}\hat{\Psi}=\hat{\mathrm{H}}\hat{\Psi}$ with the
Hamiltonian linear in derivatives. The wave function has the
so-called extra components, and the quantity
$\hat{\Psi}^{\dag}\hat{\Psi}$ has no fixed sign. The exclusion of
extra components leads to a Hamiltonian of the second order in
derivatives \cite{R1}. In addition, the PDK equations do not have
some symmetries characteristic of the Maxwell equations
(describing a massless field with spin 1) such as, for example,
the symmetry relative to the changes $\vec{E}\rightarrow\vec{H}$,
$\vec{H}\rightarrow\vec{E}$, and $t\rightarrow-t$.

In the present work, a system of first-order differential
equations for a particle with spin $S = 1$ is proposed. It is
different from the PDK equations, but its structure is similar to
those of the Maxwell and Dirac equations (with constraints). The
equations take the form
$i\hbar\partial_{t}\hat{\Psi}=\hat{\mathrm{H}}\hat{\Psi}$ with the
Hamiltonian linear in derivatives and with the wave function
without extra components, which has the meaning of the probability
amplitude.

\section{New equations of the first order in derivatives for a particle with spin 1 and with nonzero mass}

Let us recall that if the Klein--Gordon--Fock equation (here and
below, $\hbar=1$, and $c=1$)\footnotetext[1]{$^{)}$\,\,\,This is
an English translation of the article published \\ in Ukrainian
Journal of Physics in 1993 }
%(1)
\begin{equation}
\left(\Box-m^{2}\right)A_{\mu}=0, \label{E1}
\end{equation}
supplemented by the condition
%(2)
\begin{equation}
\partial^{\mu}A_{\mu}=0, \label{E2}
\end{equation}
is considered to be the initial equation for a vector massive
field, then the Proca equations of the first order in derivatives
follow from (\ref{E1}) and (\ref{E2}), if the antisymmetric tensor
%(3)
\begin{equation}
F_{\mu\nu}=\partial_{\mu}A_{\nu}-\partial_{\nu}A_{\mu} \label{E3}
\end{equation}
is introduced, and (\ref{E1}) is rewritten (with regard for
(\ref{E2})) as
%(4)
\begin{equation}
\partial^{\mu}F_{\mu\nu}+m^{2}A_{\nu}=0. \label{E4}
\end{equation}

Equations (\ref{E4}) together with (\ref{E3}) are the ten Proca
equations for ten components of the field (six nonzero components
$F_{\mu\nu}$ and four ones $A_{\mu}$). They can be also
represented in the matrix form (see. e.g., \cite{R2}), by using
the Duffin--Kemmer matrices.

If one introduces the notations (by analogy with electrodynamics)
%(5)
\begin{equation}
\vec{E}=-\vec{\nabla}\varphi-\frac{\partial\vec{A}}{\partial{t}},
\label{E5}
\end{equation}

%(6)
\begin{equation}
\vec{H}=\mathbf{rot}\vec{A}, \label{E6}
\end{equation}
or (what is the same)
%(7)
\begin{equation}
%F_{\mu\nu}=\left(\begin{tabular}{l c c c }
%\multicolumn{1}{c}{\,\,\,\,\,$0$\,\,} &
%\multicolumn{1}{c}{\,\,\,\,\,$E_{x}$\,\,} &
%\multicolumn{1}{c}{\,\,\,\,\,$E_{y}$\,\,}&
%\multicolumn{1}{c}{\,\,\,\,\,\,\,$E_{z}$\,\,\,}
%\\\\
%\multicolumn{1}{c}{\,\,\,$-E_{x}$\,\,} &
%\multicolumn{1}{c}{\,\,\,$0$\,\,} &
%\multicolumn{1}{c}{\,$-H_{z}$\,\,}&
%\multicolumn{1}{c}{\,\,\,\,\,\,\,$H_{y}$\,\,\,}
%\\\\
%\multicolumn{1}{c}{\,\,\,$-E_{y}$\,\,} &
%\multicolumn{1}{c}{\,\,\,\,\,$H_{z}$\,\,} &
%\multicolumn{1}{c}{\,\,\,$0$\,\,}&
%\multicolumn{1}{c}{\,\,\,$-H_{x}$\,\,\,}
%\\\\
%\multicolumn{1}{c}{\,\,\,$-E_{z}$\,\,} &
%\multicolumn{1}{c}{\,$-H_{y}$\,\,} &
%\multicolumn{1}{c}{\,\,\,\,\,$H_{x}$\,\,}&
%\multicolumn{1}{c}{\,\,\,\,\,$0$\,\,\,}
%\end{tabular}
%\right),
F_{\mu\nu}=\begin{pmatrix}
  \,\,0 & \,\,\,E_{x} & \,\,E_{y} & \,E_{z} \\\\
  -E_{x} & \,0 & -H_{z} & \,\,H_{y} \\\\
  -E_{y} & \,\,\,H_{z} & \,0 & -H_{x} \\\\
  -E_{z} & -H_{y} & \,\,\,H_{x} & 0
\end{pmatrix}, \label{E7}
\end{equation}
then Eqs. (\ref{E4}) take the form
%(8)
\begin{equation}
\frac{\partial\vec{E}}{\partial{t}}=\mathbf{rot}\vec{H}+m^{2}\vec{A},
\label{E8}
\end{equation}
%(9)
\begin{equation}
\mathbf{div}\vec{E}=-m^{2}\varphi. \label{E9}
\end{equation}

Thus, the Proca equations (\ref{E3}) and (\ref{E4}) look as
(\ref{E5}), (\ref{E6}) and (\ref{E8}), (\ref{E9}) in other
notations.

It is noteworthy that these equations have no symmetry related to
the change $\vec{E}\rightarrow\vec{H}$,
$\vec{H}\rightarrow\vec{E}$, and $t\rightarrow{-t}$ inherent to
the Maxwell equations. The mass appears in the equations as
$m^{2}$ (though it would be natural for it to appear in the first
power as the quantity with dimension of energy, together with the
time derivative). The wave function has extra components
\cite{R1}: their number is $10$ instead of $2(2S + 1)=6$. Despite
the degeneracy of the Duffin--Kemmer matrices \cite{R2}, the
system of equations (\ref{E5}), (\ref{E6}), (\ref{E8}), and
(\ref{E9}) can be presented in the form
$i\hbar\partial_{t}\hat{\Psi}=\hat{\mathrm{H}}\hat{\Psi}$
\cite{R1}, but it can be done only after the exclusion of extra
components, and the Hamiltonian will be of the second order in
derivatives.

It is possible to propose a new version \cite{R3} of the equations
for particles with spin 1 of the first order in derivatives, where
those questions do not arise. Let the field $A_{\mu}$ (generally
speaking, a complex-valued one) satisfy Eq. (\ref{E1}) and
condition (\ref{E2}). Keeping the previous notations for $\vec{E}$
and $\vec{H}$ ((\ref{E5}) and (\ref{E6})), let us introduce the
new fields $\vec{u}$ and $\vec{v}$ (it is the "third" stage of
transition from the field $A_{\mu}$ to secondary ones):
%(10)
\begin{equation}
\vec{u}=\frac{\partial\vec{H}}{\partial{t}}+im\vec{H}, \label{E10}
\end{equation}
%(11)
\begin{equation}
\vec{v}=\mathbf{rot}\vec{H}. \label{E11}
\end{equation}

Since it follows directly from Eq. (\ref{E1}) that $\vec{H}$ also
satisfies the equation
%(12)
\begin{equation}
\left(\Box-m^{2}\right)\vec{H}=0, \label{E12}
\end{equation}
and from relation (\ref{E6}) one has, as is known,
%(13)
\begin{equation}
\mathbf{div}\vec{H}=0, \label{E13}
\end{equation}
we obtain for $\vec{u}$ and $\vec{v}$:
%(14)
\begin{equation}
\left(\Box-m^{2}\right)\vec{u}=0, \label{E14}
\end{equation}
%(15)
\begin{equation}
\left(\Box-m^{2}\right)\vec{v}=0, \label{E15}
\end{equation}
%(16)
\begin{equation}
\mathbf{div}\vec{u}=0, \label{E16}
\end{equation}
%(17)
\begin{equation}
\mathbf{div}\vec{v}=0. \label{E17}
\end{equation}
Instead of the equations of the second order in derivatives
(\ref{E14}) and (\ref{E15}), we now get a system of equations of
the first order. Let us differentiate relations (\ref{E10}) and
(\ref{E11}) with respect to $t$ taking into account Eq.
(\ref{E12}) and the fact that
$\mathbf{rot}\left(\mathbf{rot}\vec{H}\right)=\vec{\nabla}\mathbf{div}\vec{H}-\triangle\vec{H}=$
(by virtue of (\ref{E13})) $=-\triangle\vec{H}$. Then, passing on
the right-hand sides from $\vec{H}$ to $\vec{u}$ and $\vec{v}$
according to (\ref{E10}) and (\ref{E11}), we get the first-order
differential equations
%(18)
\begin{equation}
\frac{\partial{\vec{u}}}{\partial{t}}=im\vec{u}-\mathbf{rot}\vec{u},
\,\,\,\,\frac{\partial{\vec{v}}}{\partial{t}}=-im\vec{v}+\mathbf{rot}\vec{v}
\label{E18}
\end{equation}
with the additional conditions (\ref{E16}) and (\ref{E17}). It is
easy to see that, conversely, Eqs. (\ref{E18}), (\ref{E16}), and
(\ref{E17}) yield the Klein--Gordon--Fock equations for each
component of the field.

First, we note that the limit $m\rightarrow 0$ in (\ref{E18}),
(\ref{E16}), and (\ref{E17}) results immediately in equations
identical, by their form, to the Maxwell equations.

Second, the equations include the mass in the first power, along
with derivatives.

Third, Eqs. (\ref{E18}), (\ref{E16}), and (\ref{E17}) do not change
their form under the transformation $\vec{u}\rightarrow \vec{v}$,
$\vec{v}\rightarrow \vec{u}$, and $t\rightarrow -t$.

Finally, the obtained equations have the form of dynamic ones,
%(19)
\begin{equation}
%i\frac{\partial{\hat{\Psi}}}{\partial{t}}=\hat{\mathcal{H}}\hat{\Psi}
%i\frac{\partial{\hat{\Psi}}}{\partial{t}}=\hat{\mathbb{H}}\hat{\Psi}
i\frac{\partial{\hat{\Psi}}}{\partial{t}}=\hat{\mathrm{H}}\hat{\Psi}
\label{E19}
\end{equation}
with the Hamiltonian linear in derivatives and with the additional
conditions (\ref{E16}) and (\ref{E17}). The wave function has the
meaning of a probability amplitude (see below).

To write the Hamiltonian of a particle in an explicit form, it is
convenient to give Eqs. (\ref{E18}) in the matrix form reminding
the Dirac equations.

\section{Comparison with the Dirac equations}
Let us introduce a 6-component wave function
%(20)
\begin{equation}
\hat{\Psi}=\frac{1}{\sqrt{2}}
\begin{pmatrix}
  u_{x} \\
  u_{y} \\
  u_{z} \\
  v_{x} \\
  v_{y} \\
  v_{z}
\end{pmatrix}, \label{E20}
\end{equation}
and the matrices
%(21)
\begin{equation}
\hat{a}_{k}=\hat{\sigma}_{2}\otimes\hat{S}_{k}, \,\,\,k=1,\,2,\,3,
\label{E21}
\end{equation}
%(22)
\begin{equation}
\hat{b}=\hat{\sigma}_{3}\otimes\hat{I}. \label{E22}
\end{equation}
Here, $\hat{\sigma}_{2}=\begin{pmatrix}
  0 & -i\\i & 0
\end{pmatrix}$ and $\hat{\sigma}_{3}=\begin{pmatrix}
  1 & 0\\0 & -1
\end{pmatrix}$ are the corresponding Pauli matrices, and $\hat{S}_{k}$ are the
matrices for spin $1$:
%(23)
\[
\hat{S}_{1}=\begin{pmatrix}
  0 & 0 & 0 \\
  0 & 0 & -i \\
  0 & i & 0
\end{pmatrix},\,\,\,\,\,\,\,\,
\hat{S}_{2}=\begin{pmatrix}
  0 & 0 & i \\
  0 & 0 & 0 \\
  -i & 0 & 0
\end{pmatrix},
\]
\begin{equation}
\hat{S}_{3}=\begin{pmatrix}
  0 & -i & 0 \\
  i & 0 & 0 \\
  0 & 0 & 0
\end{pmatrix}.
\label{E23}
\end{equation}

It is easy to verify that, in the above notations, Eqs.
(\ref{E18}) take the form
%(24)
\begin{equation}
i\frac{\partial\hat{\Psi}}{\partial
t}=\hat{\mathrm{H}}\hat{\Psi}\equiv
\left(\left(\hat{\vec{a}}\cdot\hat{\vec
{p}}\right)+m\hat{b}\right)\hat{\Psi}. \label{E24}
\end{equation}
The Dirac equations differ from Eqs. (\ref{E24}) by the change of
matrices
%(25)
\begin{equation}
\hat{a}_{k}\rightarrow\hat{\alpha}_{k}=\hat{\sigma}_{2}\otimes\hat{\sigma}_{k},\,\,\,\,\,\,
\hat{b}\rightarrow\hat{\beta}=\hat{\sigma}_{3}\otimes\hat{I}
\label{E25}
\end{equation}
($\hat{\alpha}_{k}$ and $\hat{\beta}$ belong to one of the
representations of the Dirac matrices). The comparison of
relations (\ref{E25}) with (\ref{E21}), (\ref{E22}), and
(\ref{E23}) indicates that the difference between our equations
(\ref{E24}) and Dirac ones consists formally in the difference
between spin matrices: $\hat{S}_{k}$ for the spin $S = 1$ instead
of $\hat{\sigma}_{k}$ for the spin $S = \frac{1}{2}$, which is
natural.

However, the nonformal difference is more profound. Let us recall
that the Maxwell equations
$\partial_{t}\vec{E}=\mathbf{rot}\vec{H}$,
\,\,\,$\partial_{t}\vec{H}=-\mathbf{rot}\vec{E}$ (in the matrix
form, they look like Eqs. (\ref{E24}) at $m = 0$) are not reduced
to the equation $\Box\hat{\Psi}=0$ for each of the components, if
the additional conditions $\mathbf{div}\vec{E}=0$ and
$\mathbf{div}\vec{H}=0$ are ignored. Similarly, the
Klein--Gordon--Fock equation does not follow from Eqs. (\ref{E24})
(for $m \neq 0$), if the additional conditions imposed on the wave
function ($\mathbf{div}\vec{u}=0$ and $\mathbf{div}\vec{v}=0$) are
not taken into account. This is easily seen also from Eqs.
(\ref{E18}) with conditions (\ref{E16}) and (\ref{E17}) after the
second differentiation with respect to the time. In notations of
Eqs. (\ref{E24}), this looks as follows.

Let us differentiate Eqs. (\ref{E24}) with respect to
$i\frac{\partial}{\partial t}$:
%(26)
\[
-\frac{\partial^{2}}{\partial
t^{2}}\hat{\Psi}=\hat{\mathrm{H}}^{2}\hat{\Psi}=\left(\hat{\vec{a}}\cdot\hat{\vec{p}}\right)
\left(\hat{\vec{a}}\cdot\hat{\vec{p}}\right)\hat{\Psi}+
\]
\begin{equation}
+m\left(\left(\hat{b}\hat{\vec{a}}+\hat{\vec{a}}\hat{b}\right)\cdot\hat{\vec{p}}\right)
\hat{\Psi}+m^{2}\hat{b}^{2}\hat{\Psi}. \label{E26}
\end{equation}
By virtue of the identities
%(27)
\begin{equation}
\hat{b}^{2}=\hat{I},\,\,\,\,\,\,\hat{a}_{k}\hat{b}+\hat{b}\,\hat{a}_{k}=0,\,\,\,k=1,\,2,\,3\,\,,
\label{E27}
\end{equation}
following from definitions (\ref{E21}) -- (\ref{E23}), it remains
to require that
%(28)
\begin{equation}
\left(\hat{\vec{a}}\cdot\hat{\vec{p}}\right)
\left(\hat{\vec{a}}\cdot\hat{\vec{p}}\right)\hat{\Psi}=-\triangle\hat{\Psi},
\label{E28}
\end{equation}
and then Eq. (\ref{E26}) appear to be the Klein--Gordon--Fock
equation. However, Eq. (\ref{E28}) is valid only together with the
wave function $\hat{\Psi}$ satisfying the additional conditions
(\ref{E16}) and (\ref{E17}) for its components, but the relation
(\ref{E28}) is not valid in the operator sense. In other words, as
distinct from the Dirac equation, we have
%(29)
\begin{equation}
\left(\hat{\vec{a}}\cdot\hat{\vec{p}}\right)^{2}\neq -\triangle.
\label{E29}
\end{equation}

Thus, the example of Eq. (\ref{E24}) leads to the following
generalization of the Dirac method to obtain the equations of the
first order in derivatives from the Klein--Gordon--Fock equation:
the expression $\hat{\mathrm{H}^{2}}\hat{\Psi}$ (\ref{E26}) must
be equal to $\left(-\triangle+m^{2}\right)\hat{\Psi}$ with the
solution included, and the solution must satisfy, if necessary,
some additional conditions. But the above relation should not hold
exceptionally as the operator equality. This leads to the more
general relations for matrices $\hat{a}_{k}$,\,\,$\hat{b}$ as
compared with the well-known ones defining the Dirac matrices.

\section{Spin operator, the total angular momentum as an integral of motion. Continuity equation for the current}

Consider the operator
%(30)
\begin{equation}
\hat{\vec{J}}=\hat{\vec{L}}+\hat{\vec{\Sigma}}
\equiv\left[\vec{r}\times\hat{\vec{p}}\,\right]+\hat{\vec{\Sigma}},
\label{E30}
\end{equation}
where $\hat{\vec{L}}$ is the operator of orbital angular momentum,
and the operator $\hat{\vec{\Sigma}}$ is, by definition,
%(31)
\begin{equation}
\begin{array}{c}
\hat{\Sigma}_{1}=-i\left(\hat{a}_{2}\hat{a}_{3}-\hat{a}_{3}\hat{a}_{2}\right)=\hat{I}\otimes\hat{S}_{1}\,\,,\\\\
\hat{\Sigma}_{2}=-i\left(\hat{a}_{3}\hat{a}_{1}-\hat{a}_{1}\hat{a}_{3}\right)=\hat{I}\otimes\hat{S}_{2}\,\,,\\\\
\hat{\Sigma}_{3}=-i\left(\hat{a}_{1}\hat{a}_{2}-\hat{a}_{2}\hat{a}_{1}\right)=\hat{I}\otimes\hat{S}_{3}\,\,.
\end{array}
\label{E31}
\end{equation}
Using the explicit form of the Hamiltonian from Eq. (\ref{E24}),
%(32)
\begin{equation}
\hat{\mathrm{H}}=\left(\hat{\vec{a}}\cdot\hat{\vec{p}}\right)+m\hat{b},
\label{E32}
\end{equation}
one can directly verify that operator (\ref{E30}) commutes with
the Hamiltonian:
%(33)
\begin{equation}
[\hat{\vec{J}},\,\hat{\mathrm{H}}\,]=0, \label{E33}
\end{equation}
i.e., it is an integral of motion.

It is obvious that $\hat{\vec{J}}$ \,has the meaning of the
operator of total angular momentum, and $\hat{\vec{\Sigma}}$ is
the operator of spin of the particle. Since
%(34)
\begin{equation}
\left(\hat{\vec{\Sigma}}\right)^{2}=\hat{\Sigma}^{2}_{1}+\hat{\Sigma}^{2}_{2}+\hat{\Sigma}^{2}_{3}=2\hat{I},
\label{E34}
\end{equation}
and, on the other hand,
$\left(\hat{\vec{\Sigma}}\right)^{2}=S\left(S+1\right)\hat{I}$, we
find that $S = 1$, as it should be.

The equation of continuity can be obtained in usual way, by taking
the scalar products of the first and second equations (\ref{E18})
with $\vec{u}^{\,*}$ and $\vec{v}^{\,*}$, respectively, and adding
the results with the corresponding complex conjugate equalities.
We get
%(35)
\begin{equation}
\frac{\partial}{\partial t}\rho+\mathbf{div}\vec{j}=0, \label{E35}
\end{equation}
where
%(36)
\begin{equation}
\rho=\frac{1}{2}\left\{\left(\vec{u}^{\,*}\cdot\vec{u}\,\right)+
\left(\vec{v}^{\,*}\cdot\vec{v}\,\right)\right\}, \label{E36}
\end{equation}
%(37)
\begin{equation}
\vec{j}=\frac{1}{2}\left\{\left[\vec{v}^{\,*}\times\vec{u}\,
\right]+\left[\vec{v}\times\vec{u}^{\,*} \right]\right\}.
\label{E37}
\end{equation}

Here, we used the factor $1/2$ for convenience of the comparison
with electrodynamics: for $m=0$ and real field $A_{\mu}$, we get
the field energy density instead of (\ref{E36}) and the
Umov--Poynting vector instead of (\ref{E37}).

The same relations can be obtained from Eq. (\ref{E24}) in the
matrix form. Like in the case of Dirac equations, it is possible
to obtain Eq. (\ref{E35}) from Eq. (\ref{E24}) in the standard
way, with the probability density
%(38)
\begin{equation}
\rho=\hat{\Psi}^{\dag}\hat{\Psi}, \label{E38}
\end{equation}
and the probability flow density
%(39)
\begin{equation}
\vec{j}=\hat{\Psi}^{\dag}\hat{\vec{a}}\,\hat{\Psi}. \label{E39}
\end{equation}
It is easy to verify that definitions (\ref{E38}) and (\ref{E39})
coincide with (\ref{E36}) and (\ref{E37}), respectively, if
normalization (\ref{E20}) is used. (Of course, it is possible to
normalize $\rho$ and $\vec{j}$ in another way, e.g., to the charge
density and the current density, multiplying them by the quantity
proportional to the charge of an electron.) It is obvious that the
probability density (\ref{E36}) (or (\ref{E38})) is
positive-definite.

\section{Some generalizations}

The equations for $\vec{u}$, $\vec{v}$ (\ref{E18}) (or, in the
matrix form, Eq. (\ref{E24})) together with the additional
conditions (\ref{E16}) and (\ref{E17}) can be postulated as the
initial (primary) ones, without mentioning the connection with the
field $\vec{H}$ given by (\ref{E10}) and (\ref{E11}).

Indeed, Eqs. (\ref{E18}), (\ref{E16}), and (\ref{E17}) are closed.
They yield the Klein--Gordon--Fock equation for each component of
the wave function. An analogous situation is characteristic of
components of the wave function of the Dirac equations. Therefore,
the "interpretation" of the wave function components (\ref{E10})
and (\ref{E11}) is not necessary. Moreover, other variants of the
transition from (\ref{E1}) and (\ref{E2}) to (\ref{E18}),
(\ref{E16}), and (\ref{E17}) are possible. For example, we can
verify that the new quantities
%(40)
\[
\tilde{\vec{u}}=\mathbf{rot}\vec{E},
\]
\begin{equation}
\tilde{\vec{v}}=\partial_{t}\vec{E}-im\vec{E}+m^{2}e^{-imt}\partial_{t}^{-1}e^{imt}\,\vec{\nabla}\varphi
\label{E40}
\end{equation}
satisfy the same equations (\ref{E18}) with conditions (\ref{E16})
and (\ref{E17}). (Here, the symbol\, $\partial_{t}^{-1}$ means the
integration, for which one has
$\partial_{t}^{-1}\partial_{t}\chi=\partial_{t}\partial_{t}^{-1}\chi=\chi$.
In particular, if the field \,$\chi$\, decreases rapidly enough at
$t\rightarrow-\infty$, we may assume that
$\partial_{t}^{-1}\chi\left(t\right)\equiv\begin{array}{c}
  \,\,\,\,_{t} \\
  \,\,\int \\
  ^{-\infty}
\end{array}\chi\left(\tau\right)d\tau$.) It is clear that,
instead of (\ref{E40}) or (\ref{E10}) and (\ref{E11}), it is
possible to take linear combinations of these definitions and to
obtain again Eqs. (\ref{E18}), (\ref{E16}), and (\ref{E17}).

Another variant: instead of the two-stage transition
($A_{\mu}\rightarrow\,\vec{E},\,\vec{H}\rightarrow\,\vec{u},\,\vec{v}$),
we can introduce
%(41)
\[
\tilde{\vec{H}}=\mathbf{rot}\vec{A},
\]
\begin{equation}
\tilde{\vec{E}}=-\vec{\nabla}\varphi-\partial_{t}\vec{A}+im\left(\vec{A}+e^{-imt}\partial_{t}^{-1}e^{imt}\,\vec{\nabla}\varphi\right)
\label{E41}
\end{equation}
and verify that the same equations (\ref{E18}) with conditions
(\ref{E16}) and (\ref{E17}) are obtained (where $\tilde{\vec{H}}$
will be present instead of $\vec{u}$, as well as $\tilde{\vec{E}}$
instead of $\vec{v}$\,). In this case, the transition to
electrodynamics ($m\rightarrow 0$) becomes very simple already at
the level of definitions (\ref{E41}).

At the same time, it is worth to pay attention to the fact that
the relativistic covariance of Eqs. (\ref{E18}) with additional
conditions (\ref{E16}) and (\ref{E17}) is satisfied automatically
in the case of definitions like (\ref{E10}) and (\ref{E11}), where
we start from relativistic equations (\ref{E3}) and (\ref{E4}).

We note also that the components of the wave function can be
replaced by their linear combinations. Though Eqs. (\ref{E18})
with conditions (\ref{E16}) and (\ref{E17}) will be trivially
changed in this case, this corresponds only to the transition to a
new representation of the matrices in Eq. (\ref{E24}).

\section{Equations for a particle in an external electromagnetic field}

Let the particle possess a charge $e$. As known, the natural way
to introduce the interaction of a charged particle with an
external electromagnetic field
$\mathcal{A}_{\mu}=~(\Phi,\vec{\mathcal{A}}\,)$ is the "extension"
of derivatives
%(42)
\begin{equation}
\partial_{\mu}\rightarrow\partial_{\mu}-ie\mathcal{A}_{\mu}
\label{E42}
\end{equation}
in the equations of motion and in the additional conditions. In
particular, if the definition
%(43)
\begin{equation}
\hat{\mathrm{H}}_{\mathcal{A}}\equiv\left(\hat{\vec{a}}\cdot\left(\hat{\vec{p}}-e\vec{\mathcal{A}}\right)\right)+\hat{b}m
\label{E43}
\end{equation}
is introduced, then Eq. (\ref{E24}) transforms into
%(44)
\begin{equation}
\left(i\frac{\partial}{\partial
t}-e\Phi\right)\hat{\Psi}=\hat{\mathrm{H}}_{\mathcal{A}}\hat{\Psi}
\label{E44}
\end{equation}
under the additional conditions
%(45)
\begin{equation}
\left(\left(\hat{\vec{p}}-e\vec{\mathcal{A}}\right)\cdot\vec{u}\right)=0,\,\,\,\,
\left(\left(\hat{\vec{p}}-e\vec{\mathcal{A}}\right)\cdot\vec{v}\right)=0\,.
\label{E45}
\end{equation}
Let us obtain an equation of the second order in derivatives
starting from Eq. (\ref{E44}). Acting by the operator
$\left(i\frac{\partial}{\partial t}-e\Phi\right)$ on both sides of
Eq. (\ref{E44}) and calculating its commutator with
$\hat{\mathrm{H}}_{\mathcal{A}}$, we get
%(46)
\[
\left(i\frac{\partial}{\partial t}-e\Phi\right)^{2}\hat{\Psi}=
\left(i\frac{\partial}{\partial
t}-e\Phi\right)\hat{\mathrm{H}}_{\mathcal{A}}\hat{\Psi}=
\]
\begin{equation}
=\hat{\mathrm{H}}_{\mathcal{A}}\left(i\frac{\partial}{\partial
t}-e\Phi\right)\hat{\Psi}+ie\left(\hat{\vec{a}}\cdot\vec{\mathcal{E}}\right)\hat{\Psi},
\label{E46}
\end{equation}
where $\vec{\mathcal{E}}=-\frac{\partial}{\partial
t}\vec{\mathcal{A}}-\vec{\nabla}\Phi$ is the vector of external
electric field intensity. Substituting
$\hat{\mathrm{H}}_{\mathcal{A}}\hat{\Psi}$ instead of
$\left(i\frac{\partial}{\partial t}-e\Phi\right)\hat{\Psi}$ into
(\ref{E46}) according to (\ref{E44}) and using the identity
%(47)
\begin{equation}
\hat{\mathrm{H}}_{\mathcal{A}}^{2}=\left(\hat{\vec{a}}\cdot\left(\hat{\vec{p}}-e\vec{\mathcal{A}}\right)\right)^{2}+m^{2}\,,
\label{E47}
\end{equation}
which is valid due to (\ref{E27}), we have
%(48)
\[
\left(i\frac{\partial}{\partial t}-e\Phi\right)^{2}\hat{\Psi}=
\]
\begin{equation}
=\left(\hat{\vec{a}}\cdot\left(\hat{\vec{p}}-
e\vec{\mathcal{A}}\right)\right)^{2}\hat{\Psi}+m^{2}\hat{\Psi}+
ie\left(\hat{\vec{a}}\cdot\vec{\mathcal{E}}\right)\hat{\Psi}
\label{E48}
\end{equation}
instead of (\ref{E46}).

We now use the following equality valid only with the wave
function satisfying the additional conditions (\ref{E45}):
%(49)
\[
\left(\hat{\vec{a}}\cdot\left(\hat{\vec{p}}-e\vec{\mathcal{A}}\right)\right)^{2}\hat{\Psi}=
\]
\begin{equation}
=\left(\hat{\vec{p}}-e\vec{\mathcal{A}}\right)^{2}\hat{\Psi}-e\left(\hat{\vec{\Sigma}}\cdot\vec{\mathcal{H}}\right)\hat{\Psi},
\label{E49}
\end{equation}
where $\vec{\mathcal{H}}=\mathbf{rot}\vec{\mathcal{A}}$\, is the
external magnetic field intensity, and $\hat{\vec{\Sigma}}$ is the
operator of spin (\ref{E31}). (For brevity, we omit the proof of
relation (\ref{E49}).)

As a result, we get the second-order equation
%(50)
\[
\left(\left(i\frac{\partial}{\partial
t}-e\Phi\right)^{2}-\left(\hat{\vec{p}}-e\vec{\mathcal{A}}\right)^{2}-m^{2}\right)\hat{\Psi}=
\]
\begin{equation}
=-e\left(\left(\hat{\vec{\Sigma}}\cdot\vec{\mathcal{H}}\right)-i\left(\hat{\vec{a}}\cdot\vec{\mathcal{E}}\right)\right)\hat{\Psi},
\label{E50}
\end{equation}
which differs from the Klein--Gordon--Fock equation (with the
"extended" derivatives (\ref{E42}) ) by terms on the right-hand
side. Relation (\ref{E50}) is very similar to the analogous
equation for a particle with spin $1/2$, though the term
proportional to a magnetic field arises in (\ref{E49}) and
(\ref{E50}) not only due to the the commutation relations of
matrices, but also due to the properties of the wave function
(\ref{E45}).

Equation (\ref{E50}) implies that a particle with spin $S = 1$
within the proposed approach has a normal magnetic moment equal to
the Bohr magneton (without regard for the electromagnetic
corrections).

\section{Conclusion}

A system of equations of the first order in derivatives, which
differs from the well-known ones, is constructed for a massive
particle with spin $S = 1$. By their structure, the equations have
much in common with the Maxwell and Dirac equations (with
constraints). The considered case generalizes the Dirac procedure
of constructing the equations of the first order in derivatives,
starting from the Klein--Gordon--Fock equation.

\bigskip

\begin{flushright}
{\footnotesize Received 17.12.92}
\end{flushright}

\vskip15mm
\rezume {Б.Є.~Гринюк} {%
РІВНЯННЯ ПЕРШОГО ПОРЯДКУ ЗА ПОХІДНИМИ ДЛЯ ЧАСТИНКИ ЗІ СПІНОМ
$S=1$} {Побудовано систему рівнянь першого порядку за похідними
для частинки зі спіном $S=1$ і ненульовою масою. На відміну від
рівнянь ПДК, система має вигляд динамічного рівняння
$i\hbar\partial_{t}\hat{\Psi}=\hat{\mathrm{H}}\hat{\Psi}$ (з
додатковою умовою) з гамільтоніаном, лінійним відносно оператора
імпульсу. Шестикомпонентна хвильова функція дає додатно визначену
густину ймовірності $\hat{\Psi}^{\dag}\hat{\Psi}\geq 0$. Система
рівнянь має багато спільного як з рівняннями Дірака, так і з
рівняннями Максвела.}

%\vskip10mm
%\rezume{Б.Е.~Гринюк}{%}{}

\end{document}